\documentclass[aps,prd,12point,twocolumn,nofootinbib,showpacs,superscriptaddress]{revtex4-1}

\usepackage{amssymb}
\usepackage{graphicx}
\usepackage{amsmath, amsthm}
\usepackage{epstopdf}
\usepackage{hyperref}

\usepackage{amsmath,amsfonts,amssymb,mathrsfs}
\usepackage{graphicx}
\usepackage{color}
\def\be{\begin{equation}}
\def\ee{\end{equation}}
\begin{document}
\title{Cosmological applications of the Brown-York 
quasilocal mass}

\author{Marianne Lapierre-L\'eonard} 
\email{mlapierre12@ubishops.ca} \affiliation{Department of 
Physics \& Astronomy, Bishop's University, 2600 College 
Street, Sherbrooke, Qu\'ebec, Canada J1M 1Z7}

\author{Valerio Faraoni} \email{vfaraoni@ubishops.ca} 
\affiliation{Department of Physics \& Astronomy and STAR 
Research Cluster, Bishop's University, 2600 College St., 
Sherbrooke, Qu\'ebec, Canada J1M~1Z7} 

\author{Fay\c{c}al Hammad} \email{fhammad@ubishops.ca} 
\affiliation{Department of Physics \& Astronomy and STAR 
Research Cluster, Bishop's University, 2600 College St., 
Sherbrooke, Qu\'ebec, Canada J1M~1Z7} 
\affiliation{Physics Department, Champlain 
College-Lennoxville, 2580 College Street, Sherbrooke,  
Qu\'ebec, Canada J1M~0C8}

\begin{abstract}

The Brown-York quasilocal energy is applied to three 
cosmological problems which have previously been studied 
with the Hawking-Hayward quasilocal energy (Newtonian 
simulations of large scale structure formation, turnaround 
radius in 
the present accelerating universe, and lensing by the 
cosmological constant). It is found that, in an 
appropriate gauge and to first order in 
the amplitude of the cosmological perturbations describing 
local structures, the Hawking-Hayward and the Brown-York 
quasilocal masses predict the same results.

\end{abstract}



\maketitle

\section{Introduction}
\label{sec:1}

In General Relativity (GR), the notion of total mass-energy 
of an asymptotically flat system (including its rest mass, 
stresses, kinetic and gravitational energy) is well 
understood and is identified with the Arnowitt-Deser-Misner 
 mass \cite{ADM}. Non-asymptotically flat spacetimes are 
more difficult to describe. The equivalence principle 
embodied in GR  makes it impossible to localize  
gravitational energy and, for non-asymptotically flat 
geometries,  one must then resort to a 
quasilocal 
definition of energy. Defining the quasilocal 
energy of a non-asymptotically flat spacetime is highly 
non-trivial and several quasilocal definitions
 have been introduced in the literature (see 
\cite{Szabados} for a review). It seems that the 
Hawking-Hayward quasilocal energy \cite{Hawking, Hayward} 
is more commonly used, while also the Brown-York definition 
\cite{BrownYork} has been popular. Generally speaking, the 
available definitions of quasilocal energy are quite formal 
and not 
practical to use. However, the mass of an astrophysical 
system is one of its most 
basic properties and, if a 
notion of quasilocal mass is to be useful in science, it 
should not remain confined to an abstract domain but 
it should be 
useful for practical calculations in  
astrophysics and cosmology. With this goal in mind, we have 
applied the Hawking-Hayward quasilocal energy to 
cosmology in previous publications \cite{Nbody, turnaround, 
lambdalensing}.  

The 
first problem studied was whether the Newtonian simulations 
of large scale structure formation in the (relatively) 
early universe, which are now very 
sophisticated,  are adequate to describe 
the physics, despite the fact that they are performed over 
a 
cube of size larger than, or comparable with, the Hubble 
radius at the epoch of structure formation (these 
simulations begin around redshift $z\sim 
100$) \cite{simulations}.  An answer was provided in 
Ref.~\cite{ChisariZaldarriaga}, in a special gauge and to 
first order in the metric perturbations, and later 
confirmed (to second order and using a 
gauge-invariant formalism) in Ref.~\cite{GreenWald}. The 
Newtonian simulations of large scale structure formation 
are ultimately correct although {\em a priori} they 
could have  been incorrect, because 
of the small peculiar velocities of the dark matter 
particles \cite{ChisariZaldarriaga}. A different approach 
to this problem  
using the 
Hawking-Hayward quasilocal energy quantifies the ``degree 
of non-Newtonianity'' of the physical system and agrees  
with the result of Refs.~\cite{ChisariZaldarriaga, 
GreenWald}. Later, numerical 
codes fully incorporating tiny relativistic effects 
\cite{relcodes} 
confirmed that Newtonian perturbations of a 
Friedmann-Lema\^{i}tre-Roberston-Walker 
(FLRW) space are 
indeed sufficient to describe the formation of large scale 
structures with high precision and that the relativistic 
corrections are very small, although worth searching for 
in future cosmological observations.

The second application of the quasilocal energy to 
cosmology was the analysis of the turnaround radius of  
large (spherical) structures in an accelerating universe 
dominated by dark energy, that is, in the present era 
\cite{Souriau, Stuchlik1, Stuchlik2, Stuchlik3, Stuchlik4, 
Mizony05, Stuchlik5, Roupasetal, Nolan2014, PT, PTT, DG1, 
DG2, DG3, DG4, DG5, DG6, TPT, Bushaetal, RomanoPRL, 
DarkUniverse, BhattaTomaras2017}. At the turnaround radius, 
a spherical shell of dust with zero radial initial velocity 
does not collapse but expands with zero acceleration 
because the attraction of the dark matter contained 
in this 
shell is balanced by the cosmological acceleration. Since 
the cosmological effects are suppressed by a factor of 
order $H^2 R^2$, where $H$ is the Hubble parameter and $R$ 
is the size of the local system, the cosmic acceleration is 
felt only by relatively large systems. The determination 
of the turnaround radius, and especially the meaning of 
``mass contained in the sphere of critical radius'' were 
clarified by applying the Hawking-Hayward quasilocal 
energy \cite{turnaround}, which reduces to the better known 
Misner-Sharp-Hernandez mass \cite{MSH} in spherical 
symmetry \cite{Haywardspherical}. The 
Misner-Sharp-Hernandez mass is widely used in relativistic 
fluid dynamics, in simulations of spherical gravitational 
collapse to black holes, and in black hole thermodynamics 
\cite{thermo} and is related to apparent horizons 
\cite{mylastbook}. Finally, the 
Hawking-Hayward/Misner-Sharp-Hernandez quasilocal mass was 
applied to the long-standing problem of whether the 
cosmological constant contributes directly to the 
deflection angle of light rays caused by a local 
gravitational lens. Two opposing viewpoints support 
opposite answers to this question and the decade-long 
debate is still open \cite{tutti}. A new approach based on 
the quasilocal energy \cite{lambdalensing} shows that the 
debate exists because of ambiguities in the concept of mass 
contained in the sphere grazed by the light rays and 
provides a definite answer to this problem. The scope is 
also extended, for arbitrary forms of dark energy can be 
included, not only the cosmological constant to which 
almost all of the 
previous literature was limited \cite{tutti}.

Some common physics underlies these three applications of 
the quasilocal energy to cosmology, namely the competition 
between local effects due to a localized mass distribution 
(which is a Newtonian-like perturbation of the underlying 
FLRW universe), and the effects of the cosmic expansion. 
The competition between local physics and cosmic expansion 
is a recurrent theme in cosmology 
\cite{CarreraGiulini}. 
These two competing effects should be compared in a 
covariant and gauge-invariant way, at least to first order 
in the amplitude of the cosmological perturbations, which 
is usually sufficient for all practical purposes. The 
approach using the quasilocal mass is particularly suited 
to this kind of problem. In fact, to first order in the 
cosmological perturbations, the Hawking-Hayward quasilocal 
mass $M_\text{HH}$ splits into two contributions: the first 
one, determined by the perturbation, is local and 
``Newtonian'', while the second contribution is purely 
cosmological. More precisely, one writes the perturbed FLRW 
line element in the conformal Newtonian gauge as 
\begin{eqnarray} 
ds^2&=& -\left( 1+2\phi \right) dt^2 +a^2(t) \left( 
1-2\phi \right) \left(dr^2 +r^2 d\Omega_{(2)}^2 \right) \nonumber\\
&&\nonumber\\
&=& a^2(\eta) \left[ -\left( 
1+2\phi \right) d\eta^2 +\left( 1-2\phi \right) \left( dr^2 
+r^2 d\Omega_{(2)}^2 \right) \right] \nonumber\\
&&\nonumber\\
&& \label{CNG}
\end{eqnarray}
where the scale factor $a(\eta)$ is a function 
of the conformal time $\eta$ (related 
to the comoving time $t$ by $dt=ad\eta$), $\phi(r) $ is a 
Newtonian potential  
describing the local perturbation, and 
$d\Omega_{(2)}^2=d\theta^2 +\sin^2 \theta \, d\varphi^2$ is 
the line element on the unit 2-sphere. 
The  two $\phi$s appearing in Eq.~(\ref{CNG}) should {\em a priori} 
be different, but they turn out to be equal to first order, as implied 
by the  fact that the  perturbative energy-momentum 
tensor is diagonal \cite{Weinberg}. 

Only a spatially flat 
FLRW universe is considered. To first order in the 
perturbations, and  following standard literature, vector 
and tensor  perturbations can be safely omitted from the 
line element~(\ref{CNG}) because the mass concentrations 
described by the perturbations have non-relativistic 
peculiar velocities \cite{Mukhanov, 
ChisariZaldarriaga, perts}. Vector and 
tensor perturbations should be included in second order 
calculations due to mode-mode coupling, but in this work 
second order effects are completely negligible. The 
first-order splitting of the quasilocal energy is 
\cite{hhconfo, Nbody, turnaround, lambdalensing}
\begin{eqnarray}
M_\text{HH} &=& ma + \frac{H^2R^3}{2} \left(1-2\phi 
\right) \simeq ma + \frac{H^2R^3}{2} \nonumber\\
&&\nonumber\\
& =& ma(t) + \frac{4\pi R^3}{3} \, \rho  \,,
\end{eqnarray}
where $ m$ is the local Newtonian mass responsible for the metric 
perturbation $\phi$ in Eq.~(\ref{CNG}) and $\rho$ is the 
cosmological density of the 
spatially flat FLRW 
background, related to the 
Hubble parameter by the Friedmann equation 
\be
H^2 =\frac{8\pi G}{3} \, \rho \label{Friedmann}
\ee
for a spatially flat FLRW universe.

The competition between local dynamics and cosmological 
expansion is described by the two contributions (local, 
$ma$, and cosmological, $H^2R^3 /2$)   to the 
quasilocal energy. For example, the critical turnaround 
radius of a large spherical structure in a FLRW universe is 
obtained when these two contributions are equal 
\cite{turnaround}.   In the discussion of whether 
Newtonian simulations of large scale structure formation 
are adequate, it was shown that the Newtonian, local, 
contribution to the Hawking-Hayward mass dominates over the 
cosmological contribution \cite{Nbody}. And, in a study of 
the direct contribution of the cosmological constant to the 
deflection angle of light rays by a localized gravitational 
lens, the splitting of the Misner-Sharp-Hernandez mass 
determines a splitting of the deflection angle into a 
contribution by the local lens plus a cosmological 
contribution---the latter vanishes to first order 
\cite{lambdalensing}.

The results of Refs.~\cite{Nbody, turnaround, 
lambdalensing} show 
that the quasilocal energy does not need to remain a formal 
concept relegated to the realm of abstract 
mathematical physics, 
but is actually useful in more practical applications. 
However, while this new approach  
allows one to draw definite conclusions on this kind of 
perturbative problem in 
which local dynamics competes with the cosmological expansion 
in the early or in the late universe, there remains a 
doubt. Refs. ~\cite{Nbody, turnaround, lambdalensing} 
use the Hawking-Hayward mass. The question 
arises naturally of whether the use of a different 
quasilocal  mass would provide different results. In the 
present article we set out to investigate this question by 
exploring the predictions of the Brown-York quasilocal mass 
\cite{BrownYork} in the same problems. 
Although our main motivation is to clarify the issue of 
choosing a quasilocal mass in view of  
cosmological applications, the problem is also interesting 
in principle.  If different quasilocal energy prescriptions 
provide different outcomes, even at the first-order which is testable 
with astronomical observations \cite{Lee}, then one would in principle 
have, 
besides computer simulations, an experimental way of discriminating 
between different definitions of quasilocal energy. For the three 
cosmological problems listed above, 
the analysis of the following sections leads to the 
conclusion 
that the Hawking-Hayward and the Brown-York quasilocal 
energies provide the same result to first order in the 
cosmological perturbations (second order differences are in 
principle possible but not testable with present and 
foreseeable technology for the cosmological problems 
described). A {\em caveat} is mandatory: since the 
Brown-York energy is 
gauge-dependent, the comparison with the Hawking-Hayward 
predictions can only be made in an appropriate gauge, {\em 
i.e.}, one in 
which the quasilocal energies of the unperturbed 
FLRW background spacetime coincide. 

The problems of the turnaround radius and of lensing by the 
cosmological constant or dark energy have been studied in 
the literature only with the simplifying assumption of  
spherical symmetry. There is no reason of principle for 
this choice, only calculational simplicity. The 
problem of the degree to which Newtonian simulations 
describe the formation of large scale structures, instead,  
has been 
studied without this assumption \cite{Nbody}. However, a simple 
spherical toy model already provides the essence of the 
physical argument and the same answer of the full 
discussion   
\cite{Nbody}. The 
analysis has been  
repeated without the assumption of spherical symmetry and 
is, by necessity, much more detailed but  
the physics  does not change \cite{Nbody}. In this article 
we restrict ourselves to spherical symmetry to illustrate 
the basic physics, leaving a more general analysis of the 
non-spherical case for future work. We use units in which 
the speed of light $c$ and Newton's constant $G$ are unity 
and we follow the notation of Ref.~\cite{Wald}.

\section{Comparison between quasilocal masses in a  
perturbed FLRW universe} 
\label{sec:2}

Let us assume a spherically symmetric geometry. In a gauge 
using the areal radius $R$ as the radial coordinate, 
\be
ds^2=-N^2(t,R)dt^2 +\frac{dR^2}{f^2(t,R)} +R^2 
d\Omega_{(2)}^2 \,,  \label{gauge}
\ee
the Brown-York quasilocal mass is given by 
\cite{BrownYorkLau, BlauRollier, YuCaldwell}
\be
M_\text{BY}=R\left( 1-f \right) \,.
\ee
The Brown-York mass is defined  with respect to a reference space which, 
in this case, is obtained for 
$N=f=1$, turning Eq.~(\ref{gauge}) into a Minkowski-like metric.  
The Hawking-Hayward mass, which reduces to the 
Misner-Sharp-Hernandez prescription in spherical symmetry 
\cite{Haywardspherical}, is
\be
M_\text{MSH}=\frac{R}{2} \left( 1-f^2 \right) \,,
\ee
as follows from the general definition \cite{MSH}
\be
1-\frac{2M_\text{MSH}}{R} =\nabla_cR \nabla^c R \,. 
\label{MSHgeneral}
\ee
The gauge~(\ref{gauge}) is used almost universally in 
the literature on  black hole 
thermodynamics, and also in Refs.~\cite{BrownYorkLau, 
BlauRollier, YuCaldwell}, in which the two quasilocal masses have been 
investigated and an attempt to interpret them physically has been made.

In general, the Brown-York mass is defined in terms of a 
$3+1$ splitting of spacetime and of the associated 3-metric 
and extrinsic curvature \cite{BrownYork}, which makes it 
clear that this quantity depends on the foliation or  
gauge chosen. By contrast, the Misner-Sharp-Hernandez mass 
defined by the scalar equation~(\ref{MSHgeneral}) and by 
the areal radius $R$ (which is a geometric, 
gauge-independent quantity), is gauge-invariant. 
Therefore, 
it is meaningless to compare these two quasilocal energy 
constructs in an arbitrary  gauge. In our situation, we 
want to compare the 
prediction of Misner-Sharp-Hernandez and Brown-York masses 
for a (spherically symmetric) perturbed FLRW universe. In 
order for the comparison to make sense, one should choose  
a gauge in which the two quasilocal masses coincide for 
a sphere in the 
unperturbed universe. This will be done in the following 
calculation, and shown explicitly at its end by taking the 
limit of zero perturbations.

Let us begin with the perturbed FLRW line 
element~(\ref{CNG}) and let us transform it to a gauge 
which uses the areal radius 
\be\
R(\eta, r)=a(\eta) r \sqrt{1-2\phi(r)} 
\label{101} 
\ee
as the radial coordinate. 
In the end, we will retain only the lowest order terms in 
the metric perturbation $\phi$, $r\phi'$, and $HR$, where 
$H=\dot{a}/a $ (an overdot denoting differentiation with 
respect to the comoving time $t$) is the Hubble parameter 
and $HR$ is approximately the size $R$ of the local 
perturbation in units of the Hubble radius $H^{-1}$. 
Equation~(\ref{101}) yields 
\be
dr=\frac{ \sqrt{1-2\phi}}{a\left( 1-2\phi -r\phi' \right)} 
\left( dR -a\dot{a}r \sqrt{1-2\phi} \, d\eta \right)
\ee
which, substituted into the line element~(\ref{CNG}), gives
\begin{eqnarray}
ds^2 &=& a^2 \left\{ 
\left[ -\left(1+2\phi \right) +\frac{ 
\left(1-2\phi \right)^2{\cal H}^2R^2}{ a^2 \left(1-2\phi 
-r\phi'\right)^2} \right] d\eta^2 \right.\nonumber\\
&&\nonumber\\
&\, & \left. + \frac{\left(1-2\phi \right)^2 dR^2}{ 
a^2 \left(1-2\phi -r\phi'\right)^2}  
-\, \frac{ 2{\cal H}R \left(1-2\phi \right)^2}{a^2 
\left(1-2\phi -r\phi'\right)^2}\,d\eta dR  \right\} 
\nonumber\\
&&\nonumber\\
&\, &  +R^2 d\Omega_{(2)}^2 \,, \label{star}
\end{eqnarray}
where 
\be
{\cal H}\equiv \frac{1}{a} \, \frac{da}{d\eta}
\ee
is the Hubble parameter of the conformal time $\eta$. Since 
we 
want to arrive at a diagonal gauge, we need to eliminate 
the time-radius cross-term by redefining the time 
coordinate, $\eta \rightarrow T$, according to 
\be
dT=\frac{1}{F} \left( d\eta +\beta dR \right) \,,
\ee
where $\beta(\eta, R)$ is a function to be determined and 
$F(\eta, R)$ is an integrating factor satisfying the 
equation
\be
\frac{ \partial}{\partial R} \left( \frac{1}{F} \right) =
\frac{ \partial}{\partial \eta} \left( \frac{\beta}{F} 
\right) 
\ee
to guarantee that $dT$ is an exact differential. The use of  
$d\eta =FdT-\beta dR$ in the line element~(\ref{star}) 
gives  
\begin{eqnarray}
ds^2 &=& \left[ -a^2\left(1+2\phi \right)+
\frac{ \left(1-2\phi \right)^2{\cal H}^2R^2 }{
\left(1-2\phi -r\phi'\right)^2} \right] 
F^2 dT^2 \nonumber\\
&&\nonumber\\
&\, & +\left\{ \left[ -a^2\left(1+2\phi \right)+
\frac{ \left(1-2\phi \right)^2{\cal H}^2R^2}{\left(1-2\phi 
-r\phi'\right)^2} \right] \beta^2 \right. \nonumber\\
&&\nonumber\\
&\, & \left.
+\frac{ \left(1-2\phi \right)^2}{ \left(1-2\phi 
-r\phi'\right)^2} 
+\frac{ 2{\cal H}R \left(1-2\phi 
\right)^2\beta}{\left(1-2\phi -r\phi'\right)^2} 
\right\} dR^2 \nonumber\\
&&\nonumber\\
&\, & -2F\left\{ \beta \left[ -a^2 \left(1+2\phi \right)
+ \frac{ \left(1-2\phi \right)^2 {\cal H}^2R^2}{
\left(1-2\phi -r\phi'\right)^2 } \right] \right. 
\nonumber\\
&&\nonumber\\
&\, & \left. +
\frac{ {\cal H}R \left(1-2\phi \right)^2}{\left(1-2\phi 
-r\phi'\right)^2} \right\} dTdR +R^2 d\Omega_{(2)}^2 
\,.
\end{eqnarray}
By setting 
\begin{eqnarray}
\beta ( \eta , R) &=& \frac{ {\cal H} R \left(1-2\phi 
\right)^2 }{ 
\left(1-2\phi -r\phi'\right)^2 A^2 } \nonumber\\
&&\nonumber\\
&= & \frac{ {\cal H} R \left(1-2\phi \right)^2}{
a^2 \left(1+2\phi\right) 
\left(1 -2\phi -r\phi'\right)^2 -{\cal H}^2 R^2    
\left(1-2\phi \right)^2 } \,,\nonumber\\
&&
\end{eqnarray}
where   
\be
A^2\left(\eta, R \right) = a^2 \left(1+2\phi \right)
- \frac{ \left(1-2\phi \right)^2{\cal H}^2R^2}{ 
\left(1-2\phi -r\phi'\right)^2} \,,
\ee
the $dT dR$ cross-term is eliminated and we are left 
with the diagonal line element
\be
ds^2=-A^2F^2dT^2 +B^2 dR^2+R^2 d\Omega_{(2)}^2 \,,
\ee
where
\be
B^2 =  \frac{ {\cal H}^2R^2 \left(1-2\phi \right)^4}{
\left(1-2\phi -r\phi'\right)^4 A^2} 
+ \frac{ \left(1-2\phi \right)^2}{\left(1-2\phi 
-r\phi'\right)^2} \,.
\ee
Algebraic manipulations bring this metric coefficient to 
the form
\be
B^2 = \frac{ a^2 \left(1-2\phi \right)^2  \left(1+2\phi 
\right)}{ a^2
\left(1+2\phi \right) \left(1-2\phi -r\phi'\right)^2
-\left(1-2\phi \right)^2 {\cal H}^2R^2 } \,,
\ee
and, to first order, we have 
\be\label{fFunction}
f^2=\frac{1}{B^2} \simeq 1-2r\phi'-H^2R^2 \left(1-2\phi 
\right) \,,
\ee
where we used the relation between comoving  and conformal 
Hubble  parameters $H={\cal H}/a$. To first order, in 
spherical symmetry, the perturbation 
potential solves the lowest order field equations, which 
reduce to the usual Poisson equation and give  
\cite{Nbody, ChisariZaldarriaga} $\phi=-m/r$, where $m$ is 
the (constant) Newtonian mass of the spherical 
perturbation. Using this 
fact  and expanding to first order, 
one obtains the Brown-York quasilocal mass
\be
M_\text{BY}=R \left(1-f \right) \simeq ma +\frac{H^2R^3}{2} 
\left(1-2\phi \right) \,.\label{BYmass}
\ee
Similarly, the Hawking-Hayward/Misner-Sharp-Hernandez mass 
is 
\be
M_\text{MSH}=\frac{R}{2} \left(1-f^2 \right) \simeq ma 
+\frac{H^2R^3}{2} 
\left(1-2\phi \right) \,.\label{MSH}
\ee
We can now take the limit $m\rightarrow 0$ to an 
unperturbed 
FLRW universe. In this limit, one obtains
\be
M_\text{BY}^{(0)} = M_\text{MSH}^{(0)} = \frac{H^2R^3}{2} 
=\frac{4\pi 
G}{3} \, \rho R^3 
\ee
using the Friedmann equation~(\ref{Friedmann}). As 
promised, the two 
quasilocal masses coincide in this limit. This is not the 
case if other gauges are used because the Brown-York mass 
is, by definition, gauge-dependent \cite{BrownYork, 
BrownYorkLau, YuCaldwell}. Therefore, the 
equality of $M_\text{BY}^{(0)} $ and $M_\text{MSH}^{(0)} $ 
would not hold in other gauges and the comparison of the 
zero and first order masses would be meaningless.

With this {\em caveat} in mind, we have obtained the result 
that, to 
first order and in the gauge chosen, the Brown-York and 
the Misner-Sharp-Hernandez masses coincide and split nicely 
(in the same way) into local and cosmological parts in a 
perturbed FLRW 
universe. Therefore, the results previously obtained for 
Newtonian 
$N$-body simulations of large scale structures 
\cite{Nbody}, for the turnaround radius \cite{turnaround}, 
and for lensing by the cosmological constant 
\cite{lambdalensing}, 
which hinge on this decomposition, will 
hold exactly in the same way as discussed for the 
Hawking-Hayward quasilocal mass.

\section{Conclusions} 
\label{sec:3}

In the context of the three cosmological problems 
considered, the Brown-York quasilocal mass provides the 
same results previously obtained with the Hawking-Hayward 
quasilocal mass \cite{Nbody, turnaround, lambdalensing}. 
This conclusion follows from the clean  
splitting of the Brown-York mass into a cosmological 
contribution  $H^2R^3/2$ (where $H$ is the
Hubble parameter and $R$ is the size of the system) and a 
local contribution $ma$ (where $m$ is the Newtonian mass 
of the local perturbation and $a(t)$ is the scale factor 
of the FLRW background), to first order. A third 
contribution $-H^2R^3 \phi$ (where $\phi$ is the 
Newtonian potential of the local perturbation) is 
completely negligible in comparison with the first two.  
This splitting is analogous to that found for the 
Hawking-Hayward or the Misner-Sharp-Hernandez mass. 
However, the Brown-York mass is a gauge-dependent quantity 
even in spherical symmetry, as is clear from its 
definition involving the extrinsic 
curvature \cite{BrownYork} and the $3+1$ 
splitting of 
spacetime into space and time which depends on the 
observer already in special relativity. By contrast,  the 
Misner-Sharp-Hernandez mass is a true scalar, 
gauge-independent,  
quantity, which is shown by its 
definition~(\ref{MSHgeneral}) \cite{Haywardspherical}.

The gauge dependence here means observer-dependence. The Brown-York energy 
depends on the observer in the sense that it depends on the particular 
spacetime foliation one chooses. As an example, it can 
easily be 
verified that this energy 
vanishes for the de Sitter metric in the comoving 
$\left( t,r \right)$-coordinates, such that 
$ds^2=-dt^2+a(t)^2\left( dr^2+r^2d\Omega_{(2)}^2  \right)$, 
where the curvature, due to the cosmological constant $\Lambda$, cannot be 
detected by those coordinates. However the energy is non-zero, for 
example, in Schwarzschild-like coordinates  such that 
\be
ds^2=-\left( 1-\frac{\Lambda r^2}{3} \right) dt^2 + \left( 
1-\frac{\Lambda r^2}{3} \right)^{-1} dr^2+ r^2d\Omega_{(2)}^2  \,,
\ee   
because the curvature due 
to the cosmological constant becomes then encoded in the spatial component 
of the metric. Thus, this energy allows one to choose from which 
perspective (foliation) one wishes to measure the gravitational energy of 
spacetime. The Brown-York energy corresponds to the energy 
of spacetime as seen by a specific observer, whereas the 
Misner-Sharp-Hernandez energy  corresponds to the 
energy encoded in the {\textit{full}} metric of that spacetime. The choice 
of the coordinates would then be dictated by the simple goal to recover 
the contribution to the energy coming from the cosmological 
constant. This is 
actually the motivation that led us to choose the gauge (\ref{101}) for 
extracting such an energy for an expanding universe.

The difference between the Brown-York and 
 the Misner-Sharp-Hernandez masses is 
analogous to the difference between the relativistic invariant 
 $m^2= -p_\mu 
p^\mu=E^2 - p^2$ and the four-vector $p^\mu=\left( E, p^i \right)$. The 
components of the 
latter depend on the observer. However, unlike this relativistic example 
where the $p^0$ component of the four-vector coincides with the 
invariant $m$ in the reference frame in which $p^i=0$, 
 the Brown-York mass 
cannot coincide with the Misner-Sharp-Hernandez mass 
within the same 
coordinate system, except in Minkowski spacetime in which both vanish. 
They can only coincide at the first order, as we found above. Therefore, 
the 
higher orders are, in this sense, very important for a more complete and a 
deeper comparison between the two quasilocal concepts. This fact is 
actually what would make one able to argue for one definition or the other 
and be able to decide which one is the most adequate for a given problem.
However, for the three cosmological problems considered, the second 
order contributions are extremely small and unobservable.

As for our comparison between the Brown-York and the Hawking-Hayward 
formulations, we have based our conclusion on the fact that, for spherical 
symmetry, Hawking-Hayward coincides at the first order with 
Misner-Sharp-Hernandez which, in turn, coincides with Brown-York for the 
specific spacetime foliation chosen for the latter. It remains true, 
however, that just as for Brown-York, the quasilocal Hawking-Hayward 
concept also depends on the 2-surface chosen, on which the various  
geometrical quantities, such 
as the induced Ricci scalar, the expansion tensor, the shear tensor, and 
the twist vector, are defined. The latter all appear as contracted 
scalars, and hence as invariants, inside the defining integral of the 
energy, but remain nonetheless surface-related quantities that depend on 
the particular 2-surface selected. Our result shows the 
coincidence of both concepts only for the particular foliation 
(\ref{101}). As such foliation is sufficient for the Brown-York energy to 
detect the cosmological contribution, we have not investigated here 
whether the two concepts would still be in agreement for other 
foliations.

As a
consequence, 
the comparison of the Brown-York and 
the Misner-Sharp-Hernandez masses and of their effects  for 
a certain spacetime (in our case, the 
perturbed post-Friedmannian space~\ref{CNG}) is, in 
general, 
meaningless. It acquires some physical meaning only when a 
gauge is found in which the Brown-York mass of the 
unperturbed  FLRW universe reduces to the 
Misner-Sharp-Hernandez mass of the same geometry. This 
gauge is  
identified as one in which the metric is explicitly 
spherically symmetric, diagonal, and uses the areal radius 
$R$ as the radial coordinate. (This constraint restricting to  spherical 
symmetric and diagonal metrics is only due to the fact that, unlike the 
Brown-York and the Hawking-Hayward  masses, 
the Misner-Sharp-Hernandez  mass is defined only for  
spherically symmetric metrics.) This gauge is the one used  
in black hole thermodynamics \cite{thermo} and was also used 
in Refs.~\cite{BrownYorkLau, BlauRollier, YuCaldwell} 
attempting to provide physical interpretations of both  
quasilocal masses. The gauge-dependence of the 
Brown-York mass should be kept in mind at all times in our 
claim that it reproduces the results obtained with the 
Misner-Sharp-Hernandez mass in \cite{Nbody, turnaround,
lambdalensing}. Indeed, this is true only in the particular 
gauge adopted and the comparison would be meaningless if 
no gauge existed in which the two results can actually be 
compared. 

Also, without knowledge of the previous results obtained with the 
Hawking-Hayward mass \cite{Nbody, turnaround, 
lambdalensing}, a parallel calculation using the 
Brown-York mass would be meaningless as it would only give again results 
which are completely dependent on the gauge choice.

That said, let us look at the interpretation of our 
result. The 
local contribution $ma$ to the Brown-York mass may look 
puzzling at first sight, but it can be understood by 
keeping in mind that, in the geometrized units used, a mass 
is also a length (in the same way that $2m$ is the 
Schwarzschild radius in the Schwarzschild solution of GR) 
and the length scale $m$ determined by the local 
perturbation becomes the (physical) comoving scale $ma$ 
when embedded in a FLRW background.

The decomposition of the Brown-York quasilocal mass 
obtained here in a suitable gauge coincides, to first order, with the 
decomposition of the Hawking-Hayward quasilocal mass 
previously obtained in Refs.~\cite{Nbody, turnaround, 
lambdalensing}. The solution of the three problems studied 
in these references (namely Newtonian large-scale structure 
formation in the early universe, the turnaround radius in 
the present accelerated universe dominated by dark energy, 
and the direct contribution of a cosmological constant or 
dark energy to light deflection by a gravitational lens) 
hinges on the splitting of the quasilocal mass into local 
and cosmological contributions. Therefore, obtaining 
exactly the same splitting (to first order) for both 
notions of quasilocal mass is sufficient to state that the 
results obtained for these problems coincide using both 
quasilocal energy notions. On the positive side, this fact 
testifies of the usefulness and power of the quasilocal 
mass in GR and encourages further exploration of its uses 
in cosmology and in astrophysics. On the negative side, it 
is not possible, at the level of current and foreseeable  experiments, 
to discriminate between the Hawking-Hayward 
and the Brown-York quasilocal energies by applying them to 
concrete problems in gravity, as it was hoped for, with  
the extra complication that the comparison of the results 
is meaningful only in a  certain gauge due to the 
gauge-dependence of the Brown-York mass.

Here we have restricted ourselves to spherical symmetry. 
Indeed, the problems of the turnaround radius and of 
lensing by the cosmological constant have, thus far, been 
studied only in spherical symmetry \cite{tutti, Souriau, 
Stuchlik1, Stuchlik2, Stuchlik3, Stuchlik4, Mizony05, 
Stuchlik5, Roupasetal, Nolan2014, PT, PTT, DG1, DG2, DG3, 
DG4, DG5, DG6, TPT, Bushaetal, RomanoPRL, DarkUniverse, 
BhattaTomaras2017}. The restriction to spherical symmetry 
is certainly inappropriate to describe large-scale 
structure formation in the early universe. However, the 
physical argument provided in Ref.~\cite{Nbody} as to why 
Newtonian simulations are ultimately correct, obtained 
without assuming spherical symmetry, was essentially the 
same that was previously derived in the same reference 
using a much 
simpler spherical toy model. The structure of the 
non-spherical calculation and the underlying physics are 
naturally expected to be the same when a general ({\em 
i.e.}, non-spherical) analysis is performed using the 
Brown-York instead of the Hawking-Hayward quasilocal 
energy. A detailed non-spherical analysis, which will be reported 
elsewhere, is, however, still lacking in this sense before one could 
achieve completeness within such an investigation. 

As a final remark, we would like to mention here that, although our 
investigation has been carried out within the perturbed {\textit{flat}} 
FLRW spacetime, taking a more general perturbed FLRW metric would not 
change our conclusions about the way the various mass definitions split 
into cosmological and local contributions. The only difference is that 
they would not agree on the local part anymore.

In fact, a spatially curved FLRW metric would only display the extra factor $(1-kr^2)^{-1}$ in front of the $dr^2$ term in (\ref{CNG}), where $k$ is the usual spatial curvature parameter that takes on the possible values $-1$, $0$, or $+1$. The only effect of this extra factor, however, is the modification of (\ref{fFunction}) into $f^2\simeq (1-2r\phi')(1-kr^2)-H^2R^2 \left(1-2\phi 
\right)$. This, in turn, will transform (\ref{BYmass}) and (\ref{MSH}) into,
respectively,
\begin{equation}
M_\text{BY}\simeq R+(ma-R)\sqrt{1-\frac{kR^2}{a^2}} 
+\frac{H^2R^3}{2} 
\left(1-2\phi \right),
\end{equation}
and 
\begin{equation}
M_\text{MSH}\simeq\frac{kR^3}{2a^2}+ma\left(1-\frac{kR^2}{a^2}\right) 
+\frac{H^2R^3}{2} 
\left(1-2\phi\right).
\end{equation}
This clearly shows that the two results do indeed agree on the non-local 
cosmological contribution to the total mass but differ at the level of the 
local contributions. Furthermore, this difference is not due to the 
gauge-dependence of the Brown-York mass, as no specific gauge would make 
the two local contributions coincide.

This local contribution difference can actually easily be understood as 
stemming from the fact that the geometric definitions of the 
Misner-Sharp-Hernandez and the Brown-York masses are different. The former 
is an observer-independent measure of the ``geometric equivalent'' of the 
total mass enclosed within an areal radius $R$ \cite{Hammad} whereas the 
latter is an observer-dependent measure of the {\textit{extrinsic 
curvature}} caused by the total mass within a specific spacetime. As 
spatially flat, closed, and open universes necessarily possess different 
spatial extrinsic curvatures, the two masses can only agree in the flat 
space case.

On the other hand, the agreement the two masses display at the level of 
the cosmological contribution to the total mass stems from the fact that 
the overall Hubble expansion of space does not distinguish between the 
flat, closed, and open universes. In fact, although the Hubble parameter 
$H$ itself differs from one spacetime to another, depending on the spatial 
curvature parameter $k$ via the Friedmann equation, the expansion of space 
is everywhere the same for a given spatial curvature.

Note that here we have only based our analysis on the Brown-York and the 
Misner-Sharp-Hernandez quasi-local masses. The conclusion drawn about the 
agreement on the cosmological contribution and the disagreement on the 
local part between the two masses, however, remains valid also for the 
case of the Hawking-Hayward quasi-local mass. This might easily be seen 
from Eqs.~(32) and (38) of Ref.~\cite{Nbody}, in which the relation 
between the Misner-Sharp-Hernandez and the Hawking-Hayward masses clearly 
indicates that the two masses will agree again on the cosmological 
contribution due to spherical symmetry.

{\em Acknowledgments:} VF thanks Peter Dunsby for pointing 
out the issue addressed in this work. We thank the 
participants to the 2017 Summer Cosmology Seminar at 
Bishop's University for discussions.  V.F. and F.H. are 
grateful to the Natural Sciences and Engineering Research 
Council of Canada for financial support.

\end{document}